\documentclass[sigconf]{acmart}

\AtBeginDocument{%
  }

\acmISBN{978-1-4503-XXXX-X/18/06}

\usepackage{booktabs}
\usepackage{array} 

\newcommand{\qcrankabs}{\textbf{QCrank}}
\newcommand{\qcrank}{\textbf{\texttt{QCrank}}}

\newcommand{\qgabs}{\textbf{Q-Gear}}
\newcommand{\qg}{\textbf{\texttt{Q-Gear}}}

\newcommand{\cq}{\textbf{\texttt{Cuda-Q}}}

\newcommand{\qis}{\textbf{\texttt{Qiskit}}}
\newcommand{\pen}{\textbf{\texttt{Pennylane}}}
\newcommand{\mpich}{\textbf{\texttt{MPICH}}}
\newcommand{\podman}{\textbf{\texttt{Podman}}}
\newcommand{\np}{\textbf{\texttt{NumPy}}}
\newcommand{\hd}{\textbf{\texttt{HDF5}}}

\usepackage{placeins}
\usepackage{float}

\usepackage{comment}

\usepackage{amsmath,amsfonts,amsthm,mathtools,nicefrac}

\usepackage{listings} 
\usepackage{xcolor}   

\usepackage{algorithm}
\usepackage{algpseudocode}
\usepackage{booktabs}

\usepackage{subcaption}

\definecolor{codegray}{rgb}{0.5,0.5,0.5}
\definecolor{codepurple}{rgb}{0.58,0,0.82}
\definecolor{backcolour}{rgb}{0.95,0.95,0.92}

\usepackage{wrapfig}
\usepackage{adjustbox}

\usepackage{makecell}
\newcommand{\thickhline}{\Xhline{2.5\arrayrulewidth}}
\usepackage{caption}

\usepackage{afterpage}
\usepackage{float}

\usepackage[T1]{fontenc}
\usepackage[utf8]{inputenc}
\usepackage{pslatex}

\usepackage{breakurl}           
\usepackage{url}                
\usepackage{xcolor}             
\usepackage[]{hyperref}         
\hypersetup{                    
  colorlinks,
  linkcolor={green!80!black},
  citecolor={red!70!black},
  urlcolor={blue!70!black}
}
\usepackage[capitalize,nameinlink]{cleveref}

\usepackage{subcaption}  
\begin{document}

\title{\Large \bf Q-GEAR: Improving quantum simulation framework
}
\newcommand\addauthornote[1]{%
  \if@ACM@anonymous\else
    \g@addto@macro\addresses{\@addauthornotemark{#1}}%
  \fi}
\author{Ziqing Guo}
\additionalaffiliation{%
  \institution{NERSC, Lawrence Berkeley National Laboratory}
  \city{Berkeley}
  \state{CA}
}
\affiliation{%
  \institution{Texas Tech University}
  \state{TX}
  \country{USA}
}
\email{ziqguo@ttu.edu}

\author{Jan Balewski}
\affiliation{%
  \institution{National Energy Research Scientific Computing Center, Lawrence Berkeley National Laboratory}
  \state{CA}
  \country{USA}}
\email{balewski@lbl.gov}

\author{Ziwen Pan}
\affiliation{%
  \institution{Texas Tech University}
  \state{TX}
  \country{USA}}
\email{ziwen.pan@ttu.edu}


\begin{abstract}
Fast execution of complex quantum circuit simulations is crucial for the verification of theoretical algorithms, paving the way for their successful execution on quantum hardware. However, mainstream CPU-based platforms for circuit simulation are well established, but slower. However, the adoption of GPU platforms remains limited because different hardware architectures require specialized quantum simulation frameworks, each with distinct implementations and optimization strategies. Therefore, we introduced \qg, a software framework that transforms \qis\ quantum circuits into \cq\ kernels. By leveraging \cq\ seamless execution on GPUs, \qg\ accelerates both CPU- and GPU-based simulations by two orders of magnitude and ten times with minimal coding effort.
Furthermore, \qg\ leverages the \cq\  configuration to interconnect the memory of GPUs, allowing the execution of much larger circuits beyond the memory limit set by a single GPU or CPU node.
Additionally, we created and deployed a Podman container and a Shifter image at Perlmutter (NERSC/LBNL), both derived from an NVIDIA public image. These public NERSC containers were optimized for the Slurm job scheduler, allowing approximately 100\% utilization of up to 1,024 GPUs. 
We present various benchmarks for  \qg\ to demonstrate the efficiency of our computational paradigm. 

\end{abstract}
\acmCodeLink{https://github.com/gzquse/Q-Gear}


\maketitle
\section{Introduction}
\begin{figure}[htbp]
    \centering
    \includegraphics[width=\linewidth]{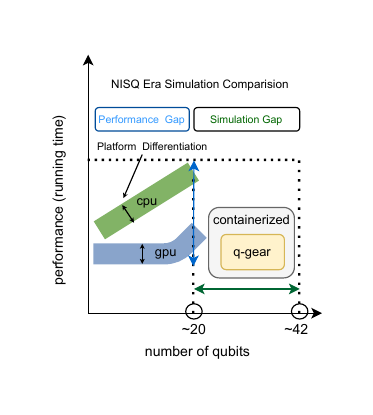}
    \Description{A diagram illustrating the challenges in quantum computing and the advantages of GPU-based quantum simulation. It highlights the scalability and performance benefits of GPU parallelization for large-scale matrix operations and the platform-agnostic capabilities of \cq\ for quantum circuit simulation.}
    \caption{The current quantum computing challenges and the general analysis of quantum simulation are introduced in \mbox{\cite{Horowitz2019progress, Gill2022taxonomy, Perdomo2018qml}}. GPU-based quantum simulation outperforms CPU simulations by overcoming performance ceilings associated with increasing qubit counts \mbox{\cite{Suzuki_2021, Kim2023cudaq};} the parallelized GPU architecture enables superior scalability and faster computation for large-scale matrix operations. \cq\ enables a platform-agnostic quantum circuit simulation by encapsulating essential simulation variables, ensuring compatibility across diverse hardware architectures. Detailed gates and state vector simulator functions are provided in \cref{app:qsvs}.}
    \label{fig:story}
\end{figure}

Quantum circuit simulation (QCS) directly models the mathematical formalism of complex quantum states. 
Quantum methods are expected to outperform classical methods in sampling and factoring problems because they bypass the need to explicitly represent and manipulate exponentially large state vectors, which is a limitation of classical algorithms. These advantages have led to the growing prominence of quantum methods in both fundamental research and widespread applications in various domains, including cryptography \cite{Pan2020wiretap}, materials science \cite{Keimer2017materials}, and quantum machine learning \cite{Biamonte2017qml}.


The most well-known approach to quantum computing systems involves designing a sequence of basic unitary operations, known as quantum gates, to transform a standard initial state into a specific target quantum state \cite{Schuld_2020}. Modern QCS tackles a diverse range of tasks, including variational quantum algorithms \cite{Cerezo2021vqa, Bittel2021np, Lubasch2020nonlinear} and hybrid quantum-classical frameworks \cite{Jeswal2019review, Altaisky2001qnn, Behrman2000simulation}. These frameworks are supported by techniques such as unitary compilation \cite{Zhang2020topological}, ansatz initialization \cite{Ostaszewski2021rl}, and circuit optimization \cite{Moro2021compiling}. The implementation of these tasks relies on tools such as \qis\ \cite{McKay2018specifications}, \pen\ \cite{pennylane}, \cq\ \cite{Kim2023cudaq}, \mpich\ \cite{gropp2002mpich2}, and \podman \cite{Stephey2022podman} containers.  
The key focus of these techniques is to improve the measurement fidelity using millions of shots and to improve the scalability of quantum computations.
\begin{figure*}[htpb]
    \centering
    \includegraphics[width=0.9\linewidth]{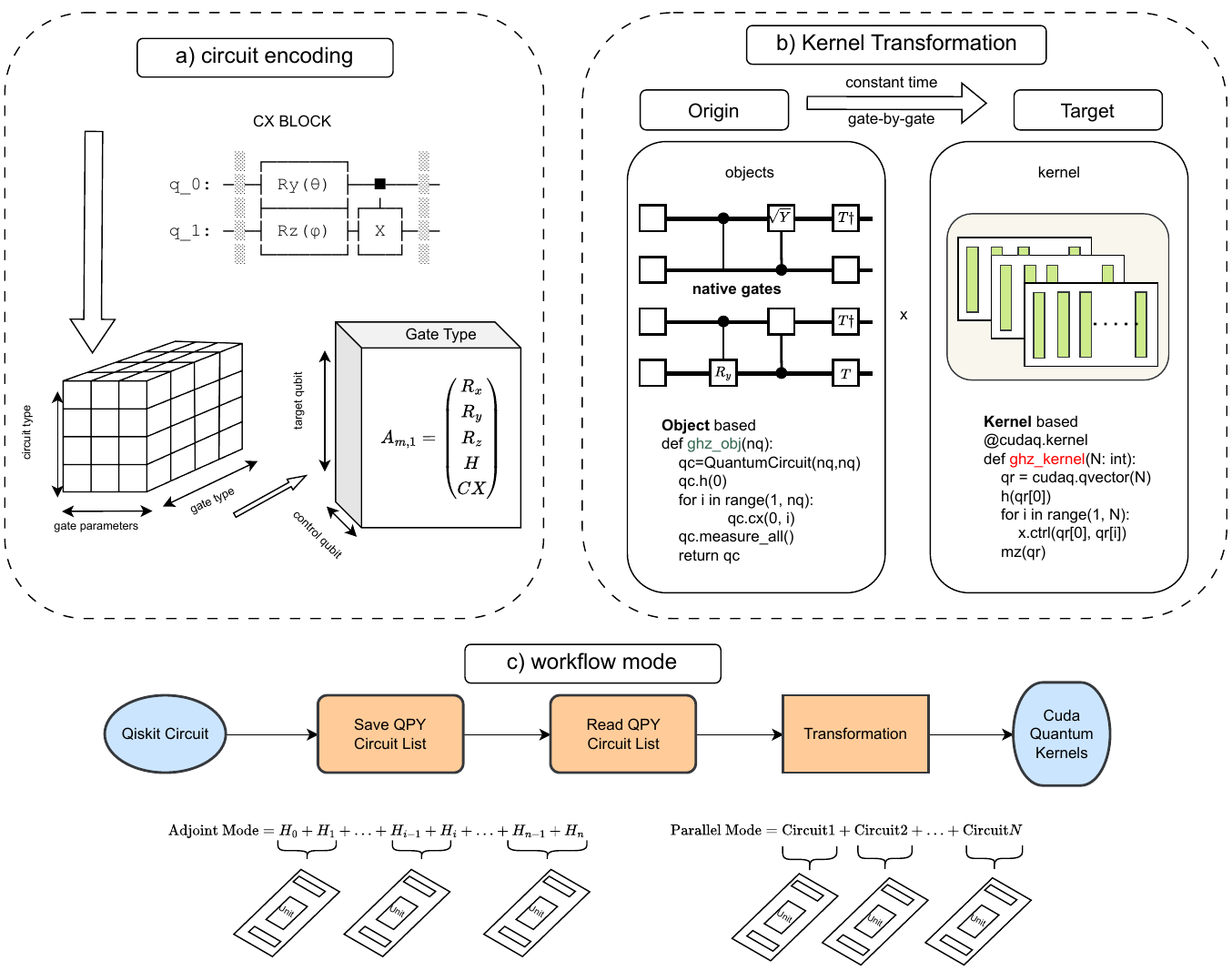}
    \Description
    {
    A summary diagram of Q-Gear methods and workflows. It illustrates three quantum circuits---Quantum Fourier Transform (QFT), Random CX Block, and QCrank---encoded as three-dimensional tensors with parameterized tensor blocks. The diagram also shows the transformation of source circuits into kernel circuits optimized for CUDA execution and highlights the workflow modes designed for large circuits and accelerated simulation.
    }
    \caption{
    Summary of Q-Gear methods and workflows. (a) Three quantum circuits---Quantum Fourier Transform (QFT), Random CX Block, and QCrank---are encoded as three-dimensional tensors, with gate types represented as parameterized tensor blocks. Each block includes \(A_{m,1}\) matrices and continuous vector-encoded gate parameters, mapping gates, control qubits, and target qubits. (b) Source circuits are transformed into kernel circuits, optimized for CUDA execution within thread warps, eliminating transformation overhead and ensuring an efficient hardware layout. (c) Workflow includes two modes designed for large circuits and accelerated simulation.
    }
    \label{fig:overview}
\end{figure*}

However, QCS efficiency remains highly sensitive to circuit layouts, with scaling challenges becoming more pronounced as circuit complexity increases. Specifically, as the number of qubits, the depth of the circuit, and the complexity of the entanglement of the circuit grow, the simulation overhead increases significantly. Furthermore, IBM experts have demonstrated a possible integration of \qis\ with a V100 GPU \cite{aer_gpu}. However, this process required installation of \qis\ from the source, a complex and non-trivial task. Their performance benchmarks on QFT circuits (up to 22 qubits) reported speedups of less than 10x utilizing single-float accuracy.


To address these QCS challenges, we introduce \qg\ that provides a platform-agnostic containerized running mode that fully utilizes the state-of-the-art GPU to accelerate the entire program, and therefore with minimal coding effort, \qg\ allows QCS of circuits that are significantly larger and faster than what is feasible on typical CPUs as shown in \cref{fig:story}.
Specifically, \qg\ optimizes performance by distributing workloads from native objects to CUDA kernels using the MPI framework. This enables scalability to higher qubit counts, reduces simulation runtime, and enhances efficiency while maintaining flexibility for user adaptability.

\section{Methods}
In this section, we present the comprehensive methods of \qg\ to encode circuit sequences into \cq\ kernels, and the architecture of the containerized workflow that fully utilizes GPU power using \podman\ \cite{Stephey2022podman} and Slurm jobs \cite{Jette2023slurm}. Detailed instructions for reproducing the pipeline are provided in \cref{app:code}, including methods for single GPU transformations, adjoint GPU modes, and cross-node configurations.

\subsection{Circuit encoding}
\label{intro:encode}
To accurately map quantum circuits, we converted the saved gate list into a three-dimensional tensor comprising matrices and tensors. As shown in (a) of \cref{fig:overview}, the first dimension encodes the type of circuit, indices of the qubits, and gate count.
The second dimension stores gate categories, control qubit indices, and target qubit indices. The third dimension captures unified gate parameters extracted from the QPY file \cite{Javadi2024qiskit}, transpiled from native gate sets, and maps the variables into predefined tensors that remain fixed but are dynamically updated based on the input quantum circuits during encoding.

The lengths of the tensors correspond to the number of circuits encoded. Although the illustrated graph demonstrates a CX block, more complex block encoding and highly entangled scenarios \cite{Camps2022fable, Mottonen2004circuits, Zhang2020topological} require larger data storage for transformation encoding.
To address this, we used the HDF5 \cite{HDF5} file format, which efficiently encodes and manages high-dimensional datasets, supports various data types (including metadata integration), and provides scalable storage for complex scientific and computational workflows. This approach ensures a constant circuit conversion time. Details and proof of the encoding process are provided in \cref{app:ce}.

\subsection{Kernel transformation}
\label{met:kernel}
To transform circuits into CUDA kernels efficiently, \qg\ directly converts \qis\ quantum circuits into GPU executable kernels. Each kernel operates as a user-controllable thread, leveraging GPU parallelism to achieve a fourfold speed-up within a single node. Untransformed Qiskit circuits are targeted for conversion into CUDA kernels, which represent transpiled pulse-like gates constrained by native QPU specifications or high-level objects containing native circuit data, as shown in \cref{fig:overview}b. 
To map object-based circuits into kernel-based representations, we define a custom CUDA quantum kernel that incorporates qubits, rotation angles, and unitary operations, thereby enabling the decoding of transformed quantum circuits directly into CUDA kernels. In parallel, this transformation fully utilizes MPI parallel memory sharing and public-channel communication. Consequently, parameterized kernel transformations preserve the structure of the final converted circuits while maximizing the computational efficiency.

\begin{figure}[htbp]
    \centering
    \includegraphics[width=0.8\linewidth]{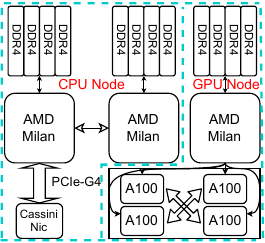}
    \Description{Diagram showing the layouts of CPU and GPU nodes. The CPU node includes two AMD EPYC 7763 processors with 64 cores each, while the GPU node includes one AMD EPYC 7763 processor paired with four NVIDIA A100 GPUs, interconnected via NVLinks and PCIe.}
    \caption{CPU and GPU node layouts.}
    \label{med:nodes}
\end{figure}

\subsection{Benchmarking hardware}
\label{med:hardware}
Based on the experimental and empirical results, we evaluated \qg\ using two hardware configurations: one with AMD EPYC 7763 processors (CPU node) and the other with NVIDIA A100 GPUs (GPU node) \cite{amd, nvidia}. 
The CPU node was powered by two AMD EPYC 7763 (Milan) processors, each with 64 cores and AVX2 support, delivering 39.2 GFLOPS per core (2.51 TFLOPS per socket). It features 512 GB of DDR4 memory with a bandwidth of 204.8 GB / s per CPU and is connected via an HPE Slingshot 11 NIC over PCIe 4.0, configured with four NUMA domains per socket (NPS=4).
The GPU node includes a single AMD EPYC 7763 (Milan) CPU with 64 cores paired with four NVIDIA A100 (Ampere) GPUs. Each GPU provides up to 2039 GB/s of memory bandwidth (80GB HBM2e) and is interconnected via four third-generation NVLinks, delivering 25 GB/s per direction per link. The node features a 256 GB DDR4 DRAM with a 204.8 GB/s CPU memory bandwidth, connected via PCIe 4.0, to the GPUs and four HPE Slingshot 11 NICs, ensuring high-performance data throughput and seamless integration of the GPU-CPU-NIC (see \cref{med:nodes}).

\subsection{Pipeline}
To demonstrate the generality of our transformation, we incorporated two key architectures into our framework: QFT transformation \cite{Weinstein2001qft, VanBeeumen2023qclab} and Quantum Image Representation \cite{Balewski2024quantum_parallel, le2011flexible}. Our heterogeneous workflow maximizes GPU utilization by integrating \podman\ for portable, consistent containerized simulations, and Slurm for efficient job scheduling, ensuring optimal task distribution, workload balance, and minimal idle resources. This approach achieved a near-peak GPU performance in large-scale quantum circuit simulations. 
For larger and more complex circuits, the simulation process partitions them into distinct Hamiltonians, representing the evolution of quantum systems (see (c) in \cref{fig:overview}). These Hamiltonians are distributed across multiple hardware resources, thereby enabling efficient parallelization. Additionally, the parallel mode allows the simultaneous execution of multiple smaller quantum circuits on separate GPUs, significantly enhancing the performance compared to sequential simulations \cite{Mottonen2004circuits}. Detailed examples are provided in \cref{app:pipeline}.

\section{Results}
\label{sec:results}

We show that \qg\ is a lightweight and efficient interface between \qis\ and \cq, which accelerates quantum circuit simulations when GPUs are available without requiring the recoding of quantum circuits. By leveraging the \cq\ ’nvidia’ target, \qg\ enables simulations of up to 42 qubits on a cluster of 1024 GPUs with a single circuit spread over all the GPUs. Below, we show the performance of \ qg for three representative cases: (1) random non-Clifford units, (2) quantum Fourier transform (QFT), and (3) \qcrank\ quantum circuits encoding grayscale images.

\setlength{\tabcolsep}{3.5pt}
\begin{table*}[hbtp]
\centering
\caption{\qg\ experiments conducted on real CPU/GPU NERSC HPC.}
{\small
\begin{tabular}{l|cc|c|cc} 
\midrule
{\bf Tasks} & \multicolumn{2}{c|}{Random entangled circuits} & QFT transform & \multicolumn{2}{c}{Quantum image encoding} \\
\midrule
{\bf Objective} & Speed-up & Scalability  & Precision & Speed-up & Reconstruction \\
& analysis & analysis & performance & analysis & performance \\
\midrule
Hardware & 32/64-core AMD EPYC & NVIDIA A100 & NVIDIA A100 & \multicolumn{2}{c}{64-core AMD EPYC} \\
 & NVIDIA A100 &&& \multicolumn{2}{c}{NVIDIA A100}\\
& HPE Slingshot 11 & HPE Slingshot 11& HPE Slingshot 11 & \multicolumn{2}{c}{HPE Slingshot 11}\\
Qubits & 28-34 & 42 & 16-33 & \multicolumn{2}{c}{15-25} \\
Max gate depth & 10,000 & 3,000 & 528 & \multicolumn{2}{c}{98,000} \\
Shots & 3,000 & 10,000 & 100 & \multicolumn{2}{c}{3M-98M}\\ 
Precision & fp32/fp64 & fp32 & fp32/fp64 & \multicolumn{2}{c}{fp64} \\
\midrule
Input size & 100/10k CX-block & 3,000 CX-block & 65K-8B bits & \multicolumn{2}{c}{5K-98K pixels}  \\
\bottomrule
\end{tabular}\\[5pt]
}
\label{tab:qgear_experiments} 
\end{table*}

\begin{figure*}[t]
    \centering
    \includegraphics[width=0.31\linewidth]{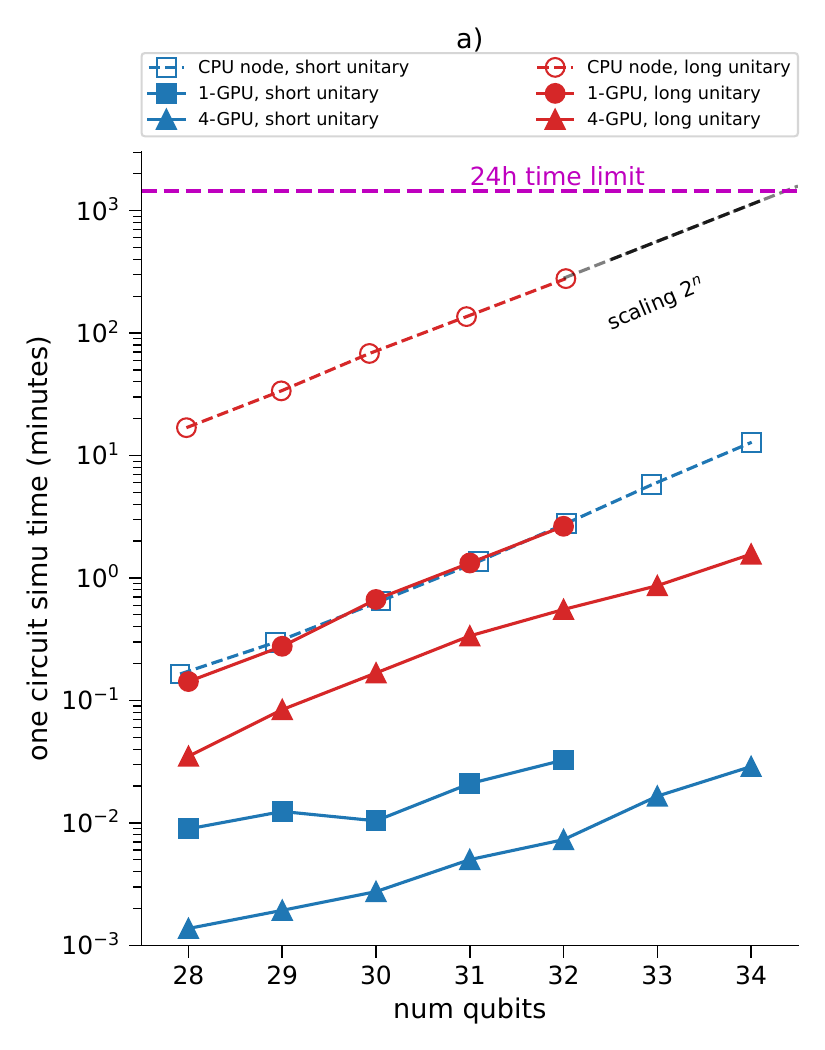}
    \includegraphics[width=0.31\linewidth]{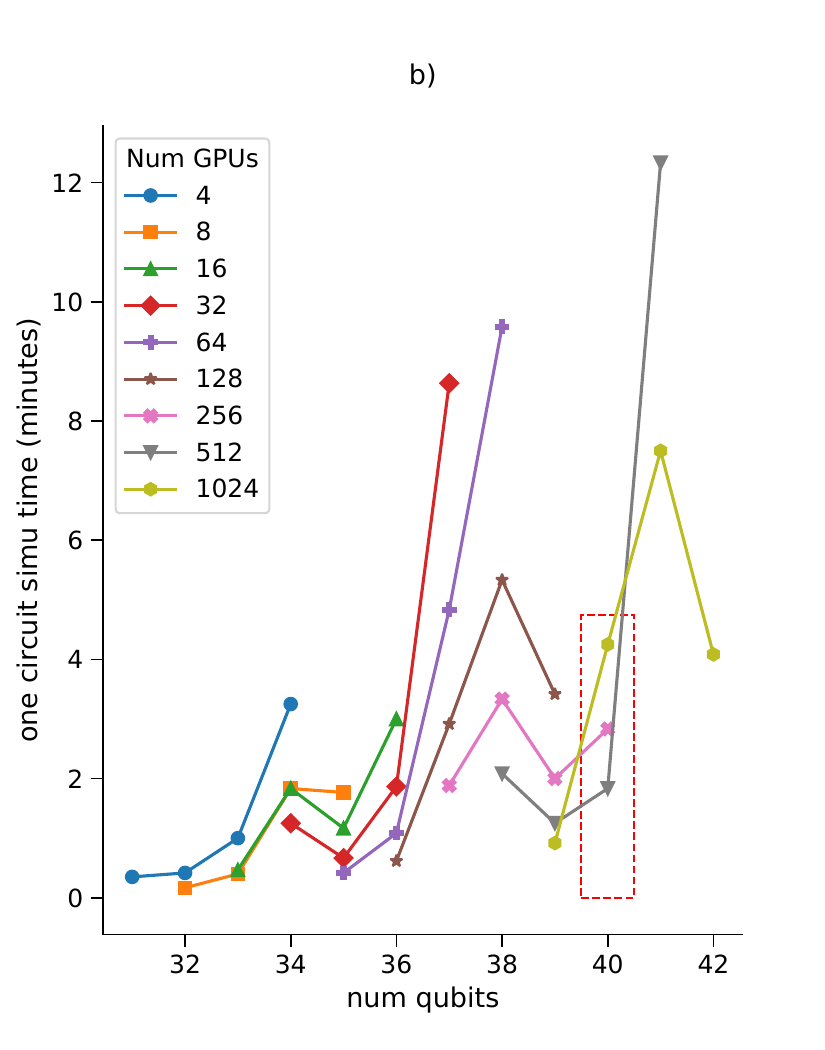}
    \includegraphics[width=0.31\linewidth]{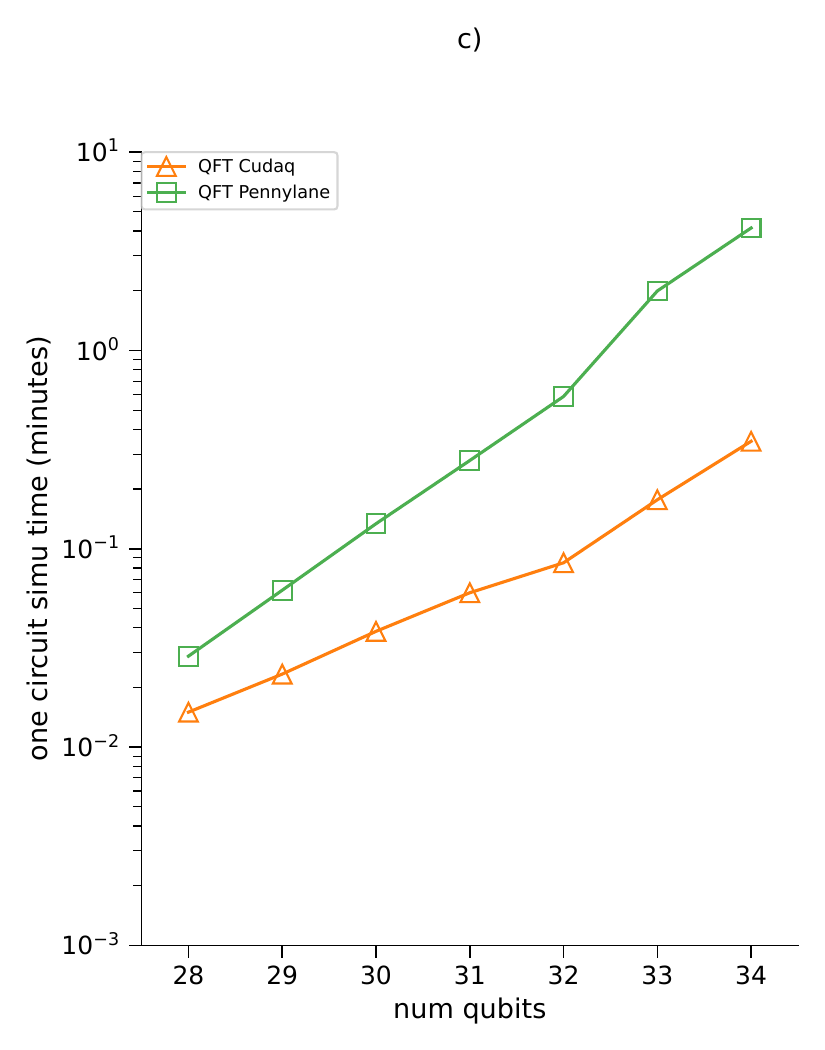}
    \caption{a) \qg\  simulation speed on a single A100 GPU and on four GPU clusters are shown as solid points. Two sizes of random unitaries, 'short' and 'long' were evaluated, as explained in the text. It is 400x faster vs. the baseline \qis\ performance on a CPU node with 128 cores.
    b) Scaling for random circuits of size 30–42 bits executed on a cluster of size 4–1024 A100 GPUs. 
    c) Scaling for  the QFT circuit execution time on four A100 GPUs with \qg\ is compared with native \pen\ execution.
    }
    \Description{Three subfigures showing: a) \qg\ simulation speed comparison on single and multi-GPU setups, b) scaling performance for random circuits on clusters of GPUs, and c) QFT circuit execution time comparison between \qg\ and native \pen.}
    \vspace{-10pt}
    \label{fig:mix}
\end{figure*}

\begin{figure}[htb]

\begin{minipage}{\linewidth}
    \centering
    \includegraphics[width=\linewidth]{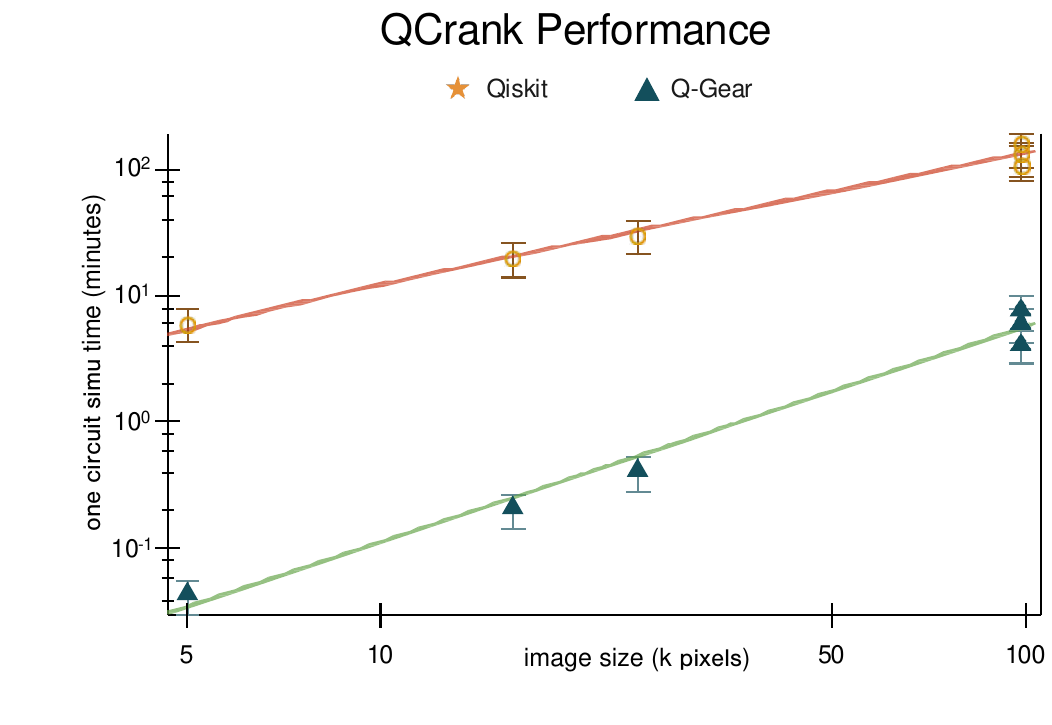}
    \caption{Performance comparison of \qis\  on CPU node and \qg\  on one A100 GPU for circuits encoding  gray-scale images  using \qcrank.  The error bars indicate the observed running time variability ($\sim$5\%). For both methods the running time scales with the image size because the pixel count is equal to the number of entangling gates in the circuit.}
    \Description{A line graph comparing the performance of \qis\ on a CPU node and \qg\ on an A100 GPU for quantum circuits encoding grayscale images. The graph shows that \qg\ achieves significantly faster execution times, with error bars indicating a 5\% variability. The running time increases with image size for both methods.}
    \vspace{-8pt}
    \label{fig:qcrank}
\end{minipage}

\vspace{30pt}

\begin{minipage}{\linewidth}
    \centering
    \captionof{table}{Quantum circuit configurations for various gray-scale images. Image dimensions determine qubit usage and shot counts for execution based on \(s*2^m\), where \(s=3000\) is the shot count per address and \(m\) is the number of address qubits.}
    \Description{This table provides quantum circuit configurations for encoding gray-scale images. It includes image dimensions, number of gray pixels, address qubits, data qubits, and shot counts. The shot count is calculated as \(s*2^m\), where \(s=3000\) and \(m\) is the number of address qubits.}
    \vspace{-10pt}
    \adjustbox{max width=0.99\linewidth}{
    \setlength{\tabcolsep}{6pt} 
    \def\arraystretch{1.25}
    \begin{tabular}{lccccc}
        \thickhline
        \textbf{Image} & \textbf{Dimensions}  & \textbf{Gray Pixels } & \textbf{Address Qubits} & \textbf{Data Qubits}  & \textbf{Shots} \\ 
        \hline
        Finger    & 64x80   &  5k & 10  & 5   & 3M  \\ 
        Shoes     & 128x128  & 16k & 11  & 8  & 6M  \\ 
        Building  & 192x128  & 25k & 12  & 6   & 12M \\ 
        Zebra     & 384x256  &98k & 13  & 12 & 24M \\ 
        Zebra     & 384x256   &98k  & 14  & 6  & 49M \\ 
        Zebra     & 384x256   &98k  & 15  & 3  & 98M \\ 
        \thickhline
    \end{tabular}
    }
    \vspace{-10pt}
    \label{tab:qc_image_data}
\end{minipage}
\end{figure}

\subsection*{Q-Gear simulation of  non-Clifford unitaries}
\label{sec:se}
To demonstrate the versatility of \qg, we compare the simulation times of randomly generated non-Clifford \cite{cross2016scalable} unitaries for a preset number of qubits. The two types of random unitaries used in this benchmark are constructed from 100 or 10,000 two-qubit blocks, where each block consists of two random single-qubit rotations followed by an entangling gate (see the visualization in \cref{fig:overview} and detailed implementation in \cref{app:cx}). This random unitary structure effectively models nontrivial workloads in quantum algorithms with increasing computational complexity.  We refer to these as 'short' or 'long' random unitaries. 

\cref{fig:mix}a demonstrates baseline performance for circuit simulation time with a CPU-based \qis\ backend Aer \cite{Javadi2024qiskit}, shown as dashed curves / open symbols. The experiments were carried out on a Perlmutter CPU node with 512 GB of DDR4 in total and 128 compute cores~\footnote{The simulations were executed on different types of hardware available at NERSC using various software backends. See \cref{med:hardware} for details on the hardware used.}.
Solid curves and closed symbols represent simulation times for the same two types of unitaries, executed using \qg\ with \cq\ on either one or four A100 GPUs, with the backend target set to 'nvidia-mgpu' \cite{Kim2023cudaq}.

For short unitaries, all available CPU RAM is exhausted at 34 qubits, shown as open squares. For long unitaries, which contain 100 times more entangling gates, the \qis\ simulation takes 100 times longer and is shown as open circles. Both cases follow a similar exponential scaling of execution time $\sim 2^n$, where $n$ is the number of qubits. One may anticipate that it would take approximately 24 h to simulate a single 34-qubit unitary with 10,000 CX gates on one CPU node.

We achieved a consistent 400-fold speedup in the simulation with \qg\ on a single GPU using the container, as indicated by the solid squares in \cref{fig:mix}a. This is non-trivial because the same \qis\ circuits were exported as \np\ arrays in the format specified by the \qg\ framework and converted to \cq\ kernels (see \cref{met:kernel}), leveraging multithreading and GPU parallelism to optimize resource utilization. 
This approach enables seamless integration between \qis\ and \cq\ backends, either within a single program or by saving \np\ circuits in the format \hd\ \cite{HDF5} for use in a separate \cq\ program. The \cq\ simulation on a single A100 GPU with a RAM of 40 GB restricts the simulable unitary to a maximum of 32 qubits. 

 In our solution, the \qg\ framework can overcome single GPU RAM limitations by setting the \cq\ target to 'nvidia-mgpu' instead of 'nvidia', which effectively combines memory from multiple GPUs. \cref{fig:mix}a shows solid triangles as the execution times of the same unitaries on 4 interconnected A100 GPUs. 
 This configuration enables the simulation of up to a 34-qubit circuit, where adding only two additional qubits requires four times more memory. The 34-qubit unitary was simulated on four GPUs in 1 min, compared to 24 h for the CPU node \qis\ simulations.
 We note that the Cuda-Q target 'nvidia-mqpu' significantly improves the execution time for 32-qubit circuits by leveraging parallelism across four GPUs, effectively utilizing them as four quantum processing units, compared to single-GPU execution.

To demonstrate \qg's ability to simulate circuits with an even larger number of qubits in the Perlmutter system, we constructed an intermediate-size unitary consisting of 3,000 random entangling blocks and varied the qubit count from 30 to 42, executing them on an increasingly larger cluster of A100 GPUs, ranging from 4 to 1024. The execution times are shown in \cref{fig:mix}b, where the colors of the symbols change with the size of the GPU cluster. It is evident that \qg\ can efficiently handle such large-circuit simulations within a reasonable time of approximately 10 min, provided that a sufficient number of GPUs are available.

Interestingly, we observed that adding more GPUs did not always reduce the execution time. For example, in the highlighted region in \cref{fig:mix}b, where the number of qubits changes from 39 to 40, the trend reverses, and a cluster with 1,024 GPUs has a lower throughput compared to a cluster with 256 GPUs. The likely cause is the network configuration: Perlmutter GPUs are stored in groups located in different racks, leading to increased communication costs if the rack boundary needs to be crossed. Another potential reason is that Perlmutter jobs are assigned to different GPUs, some of which are not warmed up (resulting in lower efficiency), thereby increasing circuit simulation time.
Therefore, we note that there is an energy trade-off to achieve the best QCS performance by setting the physical hardware within a single territory.  
\cref{fig:mix}c compares the performance of \qg\ and \pen\ (using the \texttt{lightning.gpu} backend \cite{pennylane}) for QFT circuit simulations on a 4×A100 GPU cluster. \qg\ consistently outperforms \pen, achieving significantly faster runtimes and showing better scaling with increasing circuit size. 
We note that containerized \pen\ is not optimized for large-scale simulations because the native GNU-distributed interface is not fully utilized when initializing the container in the workflow.

\subsection*{\qgabs\ simulations of image encoding}
\label{sec:qcrank}

\qcrank\ encoding~\cite{Balewski2024quantum_parallel} allows a large grayscale image to be stored in the quantum state. This provides a constructive algorithm for generating a unitary image, consisting mainly of Ry rotations and CX gates, without variationally adjustable parameters. \qcrank\ offers high parallelism in the execution of the CX gate, with the count of the CX gate equal to the number of gray pixels in the input image. A distinguishing feature of \qcrank\ is that it not only defines the procedure to recover an image previously stored on a QPU, but also allows a meaningful computation of the quantum representation.

For the selected grayscale images listed in \cref{tab:qc_image_data}, we first generated \qis\ unitaries encoding the images. We ran \qis\ simulations on one CPU node as before to establish the baseline, as shown by the open circles in \ cref{fig: qcrank}. We then used \qg\ to convert \qis\ circuits to \cq\ circuits and configured \cq\ to use one A100 GPU. The resulting simulation times are indicated by solid circles.

The \qg\ interface results in simulations that are almost two orders of magnitude faster for smaller images. The speedup decreases for larger images, most likely because achieving a similar image recovery fidelity requires more shots for larger images. In such cases, the total execution time has two components of comparable duration: unitary computation and sampling shots from this unitary. For \qis\ CPU simulations, the unitary was computed independently on each core, whereas sampling was performed in parallel on all 128 CPU cores. For \cq\ simulations on a single GPU, serial sampling was performed. Therefore, for a large number of shots, a CPU node with many cores may have an advantage over one GPU.

\section{Discussion}

This work presents two key contributions to quantum circuit migration and execution on the HPC platform. First, the \qg\ framework provides seamless migration of \qis\ circuits to \cq\ kernels, allowing efficient GPU-based simulation with \cq.
Second, we built a customized Podman-HPC image with CUDA kernels and CUDA-aware MPI, which is deployed in the public repository at NERSC.
Moreover, since Docker and Podman share the same syntax, users familiar with Docker can easily adopt this Podman image without modifying their existing workflows. This compatibility ensures that the image is versatile and practical for a wide range of researchers.

For QFT circuit execution on GPUs, we found that 
\qg\ using \cq\ outperforms \pen\, despite both leveraging the cuQuantum state vector backends. The primary reason is that when \pen\ invokes the \cq\ 'statevector' backend, the simulation process takes longer because it must first transpile high-level Python representations into low-level CUDA kernels, an additional step that introduces latency. In contrast, directly mapping quantum circuits into CUDA kernels eliminates this overhead, resulting in a more efficient simulation workflow. We note that Qiskit-GPU \cite{Javadi2024qiskit} could not be tested due to specific dependencies on PyTorch version that were not supported in NERSC. Similarly, Qulacs \cite{Suzuki_2021}  lacks maintained GPU Python packages, and its complex backend installation process using binary packages deters researchers. We were unable to evaluate TensorFlow Quantum \cite{broughton2021tensorflowquantumsoftwareframework} because it relies on deprecated packages, which makes it impractical for current research needs. These limitations highlight the superior accessibility and performance provided by \qg\ for high-performance quantum simulations on GPUs. Finally, we note that \qg\ currently operates with \cq\  with NVIDIA GPUs that rely on cuQuantum libraries \cite{bayraktar2023cuquantum}~\footnote{https://developer.nvidia.com/cuquantum-sdk}, which also support AMD GPUs with the flexibility to migrate to other HPCs. 

\section*{Acknowledgements}
This research used resources of the National Energy Research
Scientific Computing Center, a DOE Office of Science User Facility
supported by the Office of Science of the U.S. Department of Energy
under Contract No. DE-AC02-05CH11231 using NERSC award
NERSC DDR-ERCAP0034486.

{\footnotesize \bibliographystyle{unsrt}
\bibliography{main}}

\appendix
\section{Quantum state vector simulation}
\label{supp:sec:intro}
\label{app:qsvs}

In this section, we detail the procedure for encoding quantum gates in the state vector simulator, which can then be used for sampling using a user-specified number of shots. Inspired by quantum simulation with the CUDA parallelizable programming model \cite{jiao2023communication}, we transformed the gates into discrete tensors and then into a sequence of kernels, as shown in \cref{fig:overview}(a).
Specifically, a state vector simulator operates by explicitly maintaining and evolving the quantum state vector, which is represented as a \( 2^n \)-dimensional complex vector for an \( n \)-qubit system. Each quantum gate corresponds to a unitary operation represented by a \( 2^n \) vector in the Hilbert space.
For \( n \) qubits, we denote state \(\psi\) \cref{eq:hilbert} as the universal state vector.
\begin{equation}  
|\psi\rangle = \sum_{i=0}^{2^n - 1} \alpha_i |i\rangle, \quad \alpha_i \in \mathbb{C}, \quad \sum_{i=0}^{2^n - 1} |\alpha_i|^2 = 1,
\label{eq:hilbert}
\end{equation}
where \( \alpha_i \) denotes the complex amplitudes.
In fact, our experiment used \( R_x \) Gate, \( R_y \) Gate, and \( CX \) Gate, allowing for a complex and real number transformation. Furthermore, applying a single-qubit gate \( U \) to the \( k \)-th qubit involves a tensor product of the identity matrices and \( U \) shown in \cref{eq:tp}.
\begin{equation}
    U_k = I^{\otimes (k-1)} \otimes U \otimes I^{\otimes (n-k)}
    \label{eq:tp}
\end{equation}
For \( n \)-qubits, applying a CX gate to the \( k \)-th control qubit and \( m \)-th target qubit involves constructing the full unitary matrix \( U_{CX} \) indicated by \cref{eq:cx}.
\begin{equation}
    U_{CX} = \text{diag}(I^{\otimes (k-1)}, I, X, I^{\otimes (n-k-m-1)}),
    \label{eq:cx}
\end{equation}
where \( X \) acts only on the target qubit when the control qubit is in the state \( |1\rangle \). 
Referring to \cref{fig:mix}a, the GPU simulation runs time scales linearly with the increase in the CX gate tensor product operation because of the parallelism of the GPU architectures. Detailedly, CUDA-enabled GPUs consist of thousands of cores organized into streaming multiprocessors (SMs), each capable of executing multiple threads concurrently. These threads are grouped into warps that execute instructions in the lockstep, as shown in \cref{med:nodes}. Therefore, \qg\ achieves the state vector simulation parallelism by distributing independent Podman containers across these threads, allowing the simultaneous execution of different quantum circuits. 

In this section, we provide an example of a qubit system. Let us consider a 3-qubit state vector 
\begin{equation}
    |\psi\rangle = \alpha_{000}|000\rangle + \alpha_{001}|001\rangle + \dots + \alpha_{111}|111\rangle.
\end{equation}
If the control qubit is \( q_0 \) (first qubit) and the target is \( q_2 \) (third qubit), we identify all basis states where \( q_0 = 1 \): \( |100\rangle, |101\rangle, |110\rangle, |111\rangle \). In these states, the swap amplitudes for \( q_2 \) are \( \alpha_{100} \leftrightarrow \alpha_{101} \) and \( \alpha_{110} \leftrightarrow \alpha_{111} \). The \(CX\) gate involves noncontiguous memory access because the amplitudes to be swapped are scattered across the state vector. Each CX gate modifies the \( 2^{n-1} \) amplitudes in the state vector with the number of operations scaled as \( O\left(2^n \cdot d\right) \), where \( d \) is the number of CX blocks. Consequently, the workflow can achieve linear speedup by distributing quantum circuits across multiple GPUs, each processing a subset of the state vector.

\section{Circuit encoding and tensor transformation}
\label{app:ce}
\begin{lemma}\label{lem:vec}
Any quantum gate acting on \( n \) qubits can be represented as a vector in \( 2^{2n} \)-dimensional space, where the vector encodes the gate unitary matrix elements.
\end{lemma}
\textbf{Proof:}  
Let \( U \) be a quantum gate acting on \( n \) qubits. Gate \( U \) is represented by a unitary matrix of size \( 2^n \times 2^n \), where each element of the matrix corresponds to a complex number. To encode this matrix as a vector, we flattened the matrix into a single-column vector of size \( 2^{2n} \), where the elements are ordered row by row (or column by column, depending on convention).

For a single-qubit gate \( U_k \) acting on the \( k \)-th qubit, the operation can be expressed by \cref{eq:tp}, where \( I \) is the identity matrix acting on the unaffected qubits and \( U_k \) is the gate operation on the \( k \)-th qubit. This directly gives us a \( 2^{2n} \) flattened vector.
Similarly, for a multiqubit gate \( U_{ij} \) (e.g., controlled gates), the unitary matrix applies transformations that depend on both the control and target qubits, as shown in \cref{eq:cx}. The matrix \( U_{ij} \) can also be flattened into a vector of size \( 2^{2n} \) because the total number of elements remains the same, and this transformation preserves all the information about the matrix, unlike the matrix product state \cite{perez2006matrix}.

\begin{lemma}\label{lem:encode}
For any quantum circuit, a fixed-size tensor can represent the gates without loss of information, provided that the tensor size \( d \) is satisfied through \cref{eq:6}.
\begin{equation}
d \geq \max\left(|G|, |C|\right),
\label{eq:6}
\end{equation}
where \( |G| \) is the total number of gates in the largest circuit and \( |C| \) is the number of circuits encoded.
\end{lemma}

\textbf{Proof:}  
By definition, the tensor dimensions in our pipeline are determined before processing, ensuring sufficient capacity to store circuit data. During the encoding step, the tensor is initialized to its maximum allowable size \( d \) and is overridden iteratively as the circuits are processed. Given the uniform encoding strategy, the tensor length scales linearly with the number of circuits, thereby ensuring that no truncation or overflow occurs. We are now ready to prove the universality theorem.

\begin{theorem}
For a quantum circuit simulation task with \( N \) qubits (or gates), the computational time \( t \) scales exponentially on a CPU but linearly on a GPU using containerized submission, i.e.,
\begin{equation}
    t_{\text{CPU}}\left(N\right) \sim O\left(2^N\right) \quad \text{and} \quad t_{\text{GPU}}\left(N\right) \sim O\left(N\right)
    \label{the:1}
\end{equation}
\end{theorem}
\textbf{Proof:}
The quantum circuit state vector for \( N \) qubits is represented in \( 2^N \)-dimensional Hilbert space. Given the sequential nature of CPUs, each gate operation requires \( O\left(2^{2N}\right) \) operations in the worst case, because of the \( 2^N \times 2^N \) matrix representation, as shown by \cref{lem:vec}. Processing quantum gates or encoding such a vector involves sequential multiplications of unitary matrices in CPUs, followed by \cref{lem:encode}. 

We denote that the \qg\ transformation exploits parallelism by simultaneously applying operations on independent vector components inside each \podman\ container. NVIDIA A100 can partition the \( 2^{2N} \)-dimensional state space into \( P \) parallel cores, where \( P \) grows proportionally to the available hardware resources, and the state vector initialization becomes trivial. Consequently, the time complexity per gate is reduced to \( O\left(P\right) \).
Given that \( P \) is large and typically scales with \( 2^{2N} \), the effective computational complexity for GPUs becomes:
\begin{equation}
    t_{\text{GPU}}\left(N\right) \sim O\left(N\right).
    \label{eq:8}
\end{equation}
Consequently, while CPU computation time grows exponentially, GPU computation time scales linearly with problem size \( N \), provided that sufficient parallel resources are available.

\section{High dimensional data management}

To enhance the efficiency of managing high-dimensional datasets, we employ the HDF5 file format, which supports the following three properties:
\begin{enumerate}
    \item \textbf{Hierarchical Data Storage:} efficient organization of tensors, circuits, and metadata.
    \item \textbf{Scalability:} seamless handling of large datasets, reducing read/write overhead.
    \item \textbf{Compression:} storage efficiency through lossless data compression.
\end{enumerate}

Using the HDF5 format, the encoding and time saving complexity is \(O\left(N \cdot T \cdot \log(T)\right)\), with the logarithmic factor accounting for the optimized indexing. Our implementation shows that for fixed tensor sizes (\(T\)), the encoding time remains nearly constant, regardless of the circuit complexity. For example, encoding \(N = 1000\) circuits with \(T = 10^6\) took ~2 min, independent of the entanglement depth or gate count. Furthermore, HDF5 compression reduced storage by up to 50\% without affecting read/write speeds, demonstrating its efficiency for large-scale quantum datasets.

\section{Datasets}

In this section, we detail the procedure for generating three types of datasets with which \qg\ can uniformly perform the transformation using the same structured gate lists.

\subsection{Random circuit generator (CX-block)}
\label{app:cx}

We define a method \texttt{generate\_random\_gateList} that generates a randomized list of quantum circuits characterized by specific gate types and configurations. Each circuit is described by the number of qubits, specific target and control qubit indices, type of gates, and input parameters for the variational unitary quantum gates \cite{Cerezo2021vqa}. More specifically, the generator function pre-allocates the CUDA memory for circuit layout, for instance, we implement a numpy two-dimensional array for gate types M = (\texttt{h}, \texttt{ry}, \texttt{rz}, \texttt{cx}, \texttt{measure}) mapping to one hot encoding for optimization shown in \cref{eq: onehot}.
\begin{equation}
\mathbf{M}^\top =
\begin{bmatrix}
1 & 0 & 0 & 0 & 0 \\
0 & 1 & 0 & 0 & 0 \\
0 & 0 & 1 & 0 & 0 \\
0 & 0 & 0 & 1 & 0 \\
0 & 0 & 0 & 0 & 1
\end{bmatrix}
\label{eq: onehot}
\end{equation}
We then implemented block circuits with a sequential fixed number of CX gates and interleaved them with randomly paired parameterized $R_y$ and $R_z$ rotations, ensuring alternating qubit pairings. The applied algorithm is shown in \cref{alg:random_cx}.

\begin{algorithm}
\caption{Randomized Quantum Circuit Generation}
\begin{algorithmic}[1]
\State \textbf{Input:} Quantum circuit $C$
\State \textbf{Let} $Q$ be the set of qubits in $C$
\For{each gate $G$ in $C$}
    \State Sample $q_c \in Q$
    \Repeat
        \State Sample $q_t \in Q$
    \Until{$q_t \ne q_c$}
    \State Assign gate parameters \(\theta \sim \mathcal{U}\left[0, 2\pi\right]\) for randomized rotation gates
\EndFor
\end{algorithmic}
\label{alg:random_cx}
\end{algorithm}

The CX block output includes structured arrays for circuit properties (\texttt{circ\_type}), gate specifications (\texttt{gate\_type}), and gate parameters (\texttt{gate\_param}), along with metadata for the circuit generation process. To facilitate the random CX block, function \texttt{ random\_qubit\_pairs} generates a specified number of random qubit pairs from a given number of qubits. For example, (1,2) means that the first control qubit targets the second qubit. This block constructs all possible ordered pairs of qubits, excluding self-pairs (i.e., pairs where a qubit is paired). The function then randomly selects $k$ pairs with replacements from the set of valid pairs returned as a 2D numpy array.

\subsection{QFT kernel generator}
\label{app:qft_cudaq}
We customized the QFT kernel starting with the Hadamard gate layer and interconnected controlled arbitrary rotation gates, namely, \(cr1\), where each gate accepts the flattened parameters transferred from the structured \qis\ circuits output tensor-like data \cite{Weinstein2001qft}. 
For a quantum register of size $n$, the kernel applies a Hadamard gate to each qubit followed by \cref{eq:cr1}.
\begin{equation}
 CR1\left(\lambda\right) =
\begin{bmatrix}
1 & 0 & 0 & 0 \\
0 & 1 & 0 & 0 \\
0 & 0 & 1 & 0 \\
0 & 0 & 0 & e^{i\lambda}
\label{eq:cr1}
\end{bmatrix}
\end{equation}
between each qubit $i$ and all subsequent qubits $j > i$, with angles decreasing as \(
\frac{2\pi}{2^{j-i+1}}.\)
This nested loop structure introduces only $O\left(n^2\right)$ complexity. We then utilized \texttt{cudaq.qview} to facilitate efficient state manipulation and execution on NERSC platforms, making it scalable for quantum simulations.
To further optimize the kernel, we specify hyperparameters (gate fusion = 5) and approximations for negligible rotation angles to reduce the execution overhead without significant loss of fidelity, ensuring the applicability of the kernel in \qg.

\subsection{QCrank circuit generator}
\label{app:qcrank_circ}
We created a utility method for generating pre-processed grayscale image data and metadata for \qcrank,\  a quantum simulation, or computation framework. This method includes image handling, metadata construction, and preparation of quantum input formats adapted for parallel processing and \qg\ kernel transformation. 

Our method normalizes grayscale images to \(\left[-1, 1\right]\) and encodes them using trigonometric transformations to satisfy the \qcrank\ input requirements. User-configurable options allow control over image types, output paths, and addressing qubits via command-line arguments. Metadata - covering image dimensions, quantum qubits, and circuit constraints — ensures GPU compatibility. The outputs are structured as \(n_{\text{addr}} \times n_{\text{data}} \times n_{\text{img}}\) tensors, with reversed addressing qubits according to \qcrank\ conventions, and are saved as \texttt{. h5} files.

\section{Pipeline Configuration}
\label{app:pipeline}

In this section, we present the details of the two deployment modes: one named the Podman-HPC container mode and Shifter scaling node mode. In addition, we describe how these modes are used in conjunction with the \qg\ transformation.

\subsection{Podman HPC Container}
\label{app:podman}
The Podman container includes essential dependencies such as \texttt{}{copy-cuda12x}, \texttt{mpi4py}, \qis,\ and \cq,\ along with libraries for visualization and state vector simulation. To create a Podman container, we use a GCC pre-installed \texttt{cu12.0} DevOps container as the base image and integrate the native NERSC Cray MPICH, which is optimized for high-performance interconnects and parallel processing on NERSC systems. 

To streamline the workflow and maximize resource utilization, we developed a robust Slurm job submission framework that seamlessly integrates MPI parallelization (\cref{fig:overview}(c)), empowering users to configure key parameters such as the number of random circuits, simulation shots, QFT circuit reverse activation, and transformation precision. We established a Shell script, built for the containerized environment, which employs a novel technique dubbed "podman wrapper" that dynamically links batch submission variables, environment parameters (e.g., MPI rank), locally generated circuits, and output directories to the containerized execution environment. By leveraging the workflow, efficient resource orchestration and precise data management were ensured throughout the simulation pipeline. Note that the performance metrics, as captured by the NSight system, demonstrate significant improvements in computational efficiency and scalability. The complete codebase and resources are available on the GitHub for further exploration and reproducibility.

\subsection{Shifter Multiple Nodes}
\label{app:shifter}
In the multi-node mode \qg, we utilize the cuda-quantum nightly to build a Shifter image and predefined essential tools (e.g., \texttt{qiskit-aer}, \texttt{h5py}, \texttt{qiskit-ibm-experiment}, etc.) in the local scratch file system, ensuring a consistent environment for all submitted jobs. 

The architecture leverages Shifter for multinode scalable CUDA-enabled quantum simulations using \cq\ within a containerized setup as illustrated by \cref{app:podman}. The difference is that workloads are distributed across different physical nodes via MPI (cray mpich), enabling the parallel execution of complex and highly entangled quantum circuits. Theoretically, the upper bound is defined by the allocated resources, with the time complexity scaling as shown in \cref{eq:multi-node}.
\begin{equation}
    t_{\text{multi-node}} \sim O\left(\frac{2^N}{P \cdot R}\right),
    \label{eq:multi-node}
\end{equation}
where \( N \) denotes the number of qubits, \( P \) denotes the number of parallel processes per node, and \( R \) denotes the number of nodes. 
We also note that the pipeline dynamically detects the GPUs available at each node, ensuring full utilization by running threads. Inter-node communication is managed through NVLink broadcasts, which facilitate shared information across tasks. Additionally, our design supports diverse workloads with different container images, enabling large-scale quantum circuit benchmarking and high-throughput quantum image processing for further research.

\subsection{Code snippets and kernel examples}
\label{app:code}

The Slurm submission script sample is as follows:
\begin{lstlisting}
# 1 CPU mode (128 physical cores, 460 GB RAM), Qiskit Aer state-vector simulator
c64_tp4: sbatch -N 1 -c 64 -C cpu --task-per-node 4; podman-hpc; mpiexec -np 4 python run.py

# 1 GPU mode: use 1 GPU (A100, 40 GB RAM/GPU), CudaQ state-vector simulator
sbatch -N 1 -n 1 -C gpu --gpus-per-task 1; podman-hpc; python run.py --target nvidia-mgpu

# 4 GPUs mode: use 4 GPUs (A100, CudaQ state-vector simulator)
sbatch -N 1 -n 4 -C gpu --gpus-per-task 1; podman-hpc; mpiexec -np 4 python run.py --target nvidia-mgpu

# 4 Nodes mode: use 16 GPUs (80 GB RAM/GPUs A100, CudaQ state-vector simulator)
sbatch -C "gpu&hbm80g" -N4 --gpus-per-task=1 shifter bash -l -c "$CMD"
\end{lstlisting}

\section{\qcrankabs\ Benchmark details}
\label{app:benchmark}

\begin{figure*}[t]           
  \centering

  \begin{subfigure}{.48\textwidth}
    \centering
    \includegraphics[width=\linewidth]{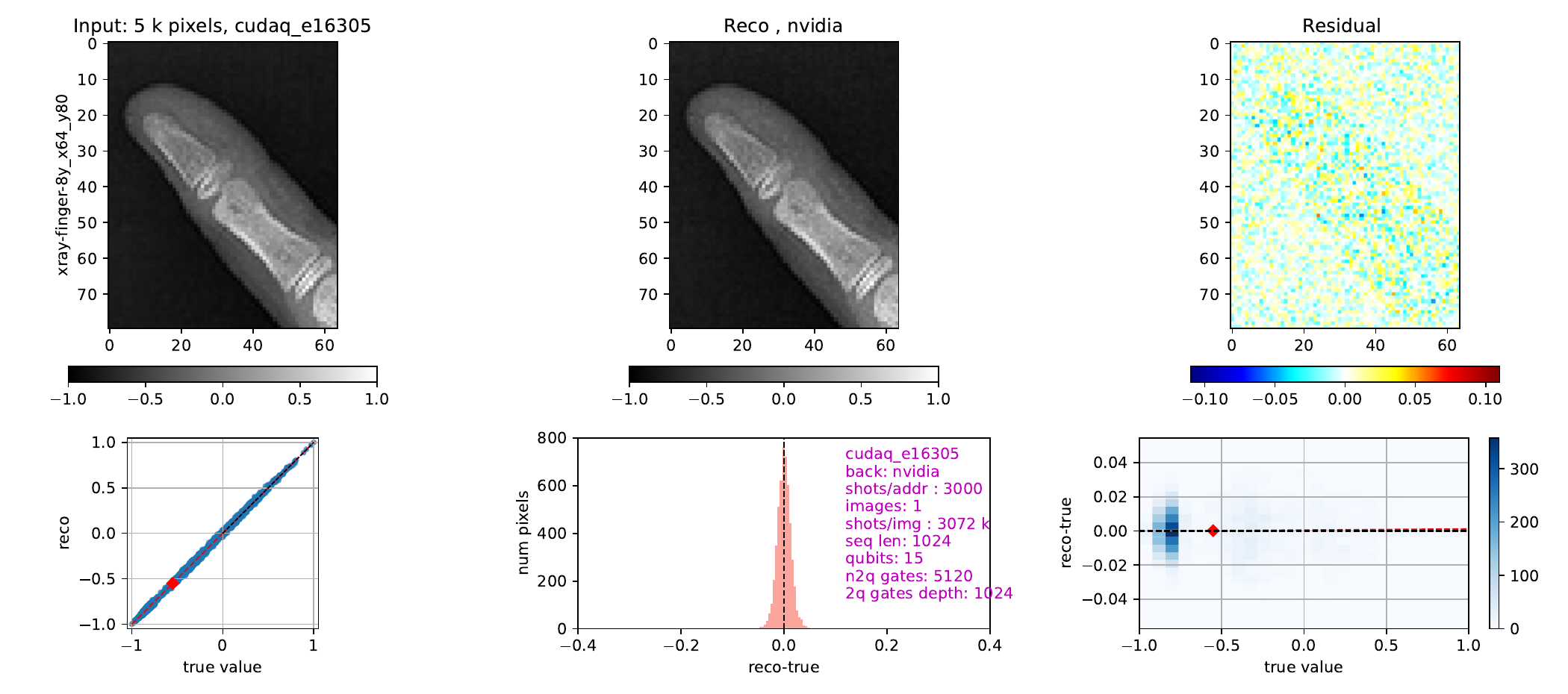}\par
    \vspace*{2pt}%
    \includegraphics[width=\linewidth]{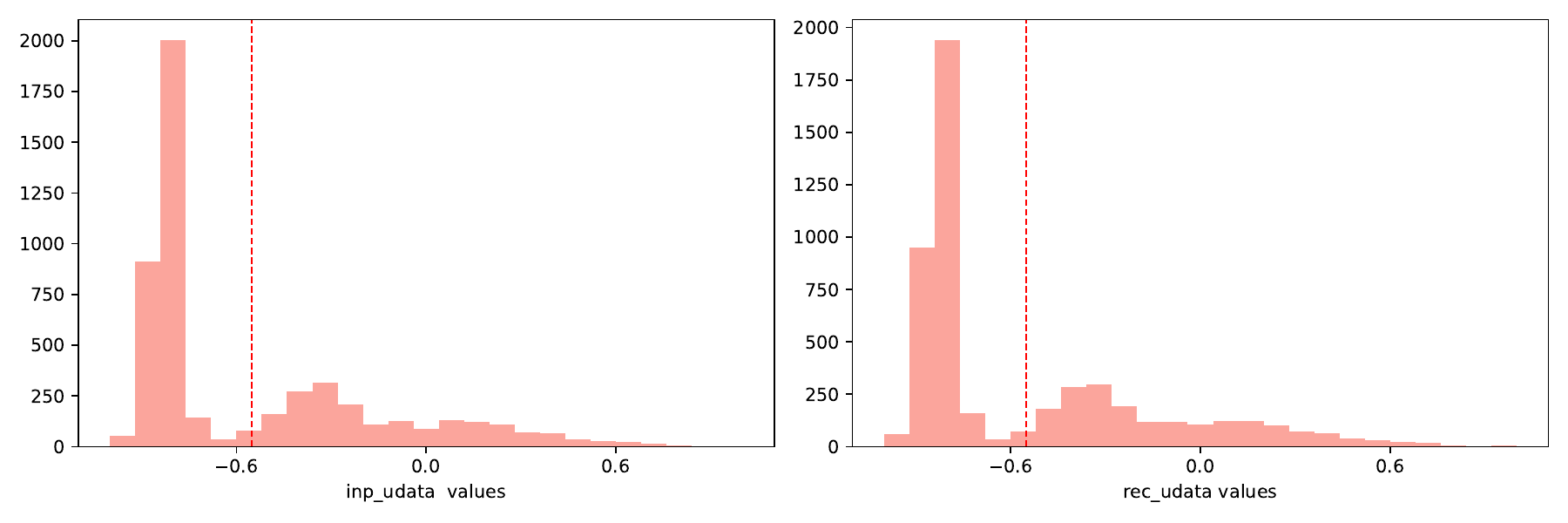}
    \caption{5 k-pixel grayscale image (Finger).}
    \label{fig:finger_combined}
  \end{subfigure}\hfill
  %
  \begin{subfigure}{.48\textwidth}
    \centering
    \includegraphics[width=\linewidth]{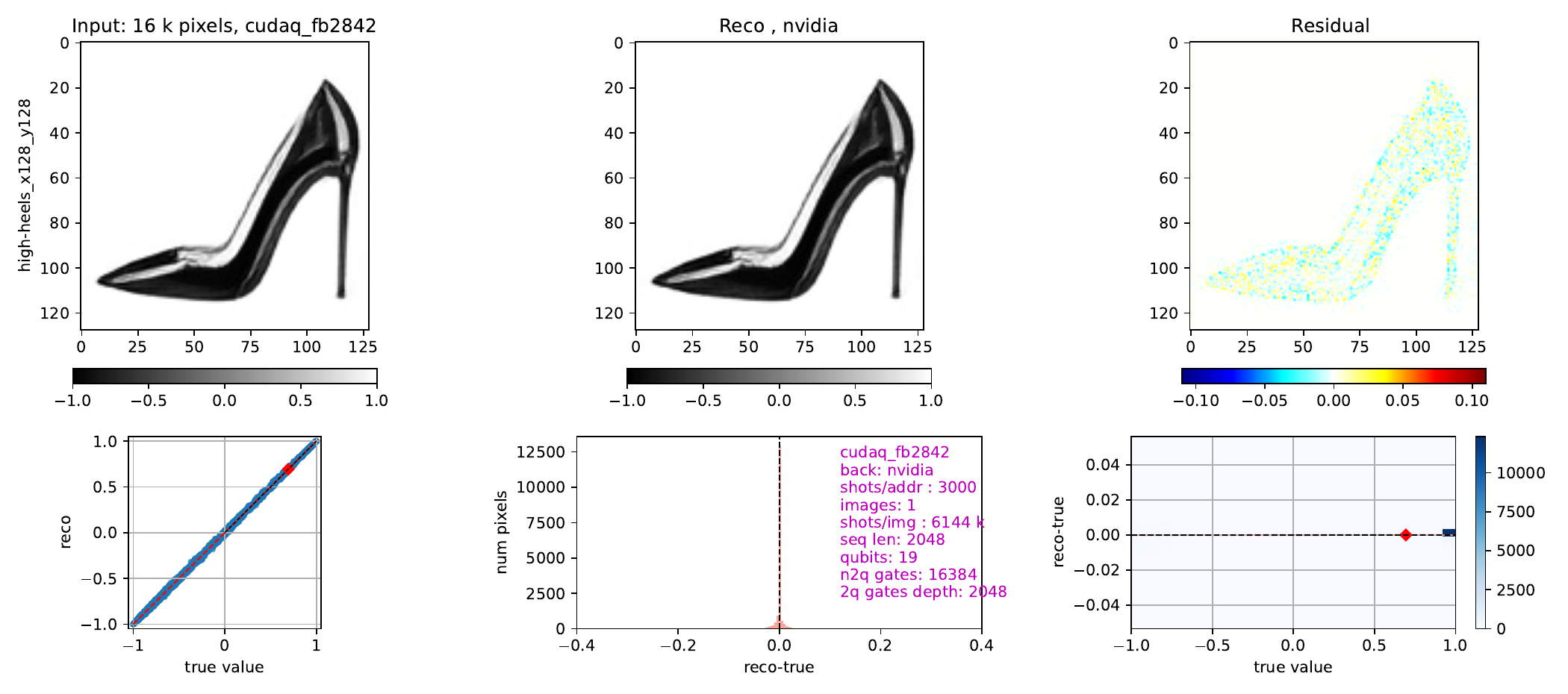}\par
    \vspace*{2pt}%
    \includegraphics[width=\linewidth]{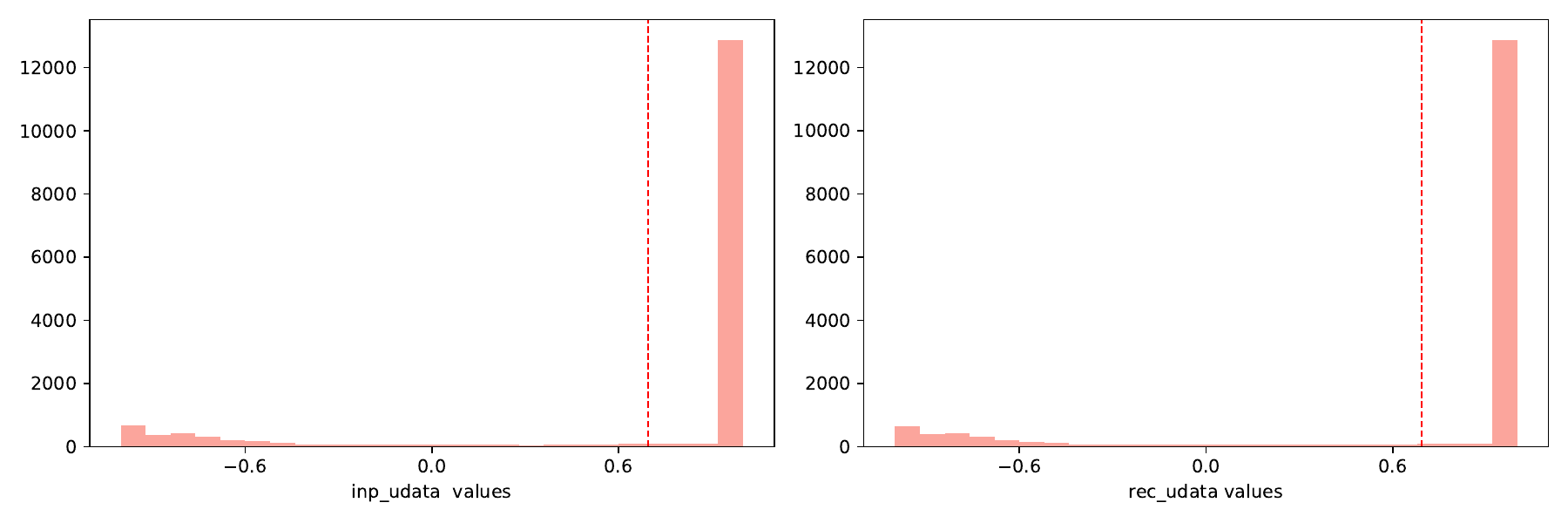}
    \caption{16 k-pixel grayscale image (Shoes).}
    \label{fig:shoes_combined}
  \end{subfigure}

  \vspace{8pt}              

  \begin{subfigure}{.48\textwidth}
    \centering
    \includegraphics[width=\linewidth]{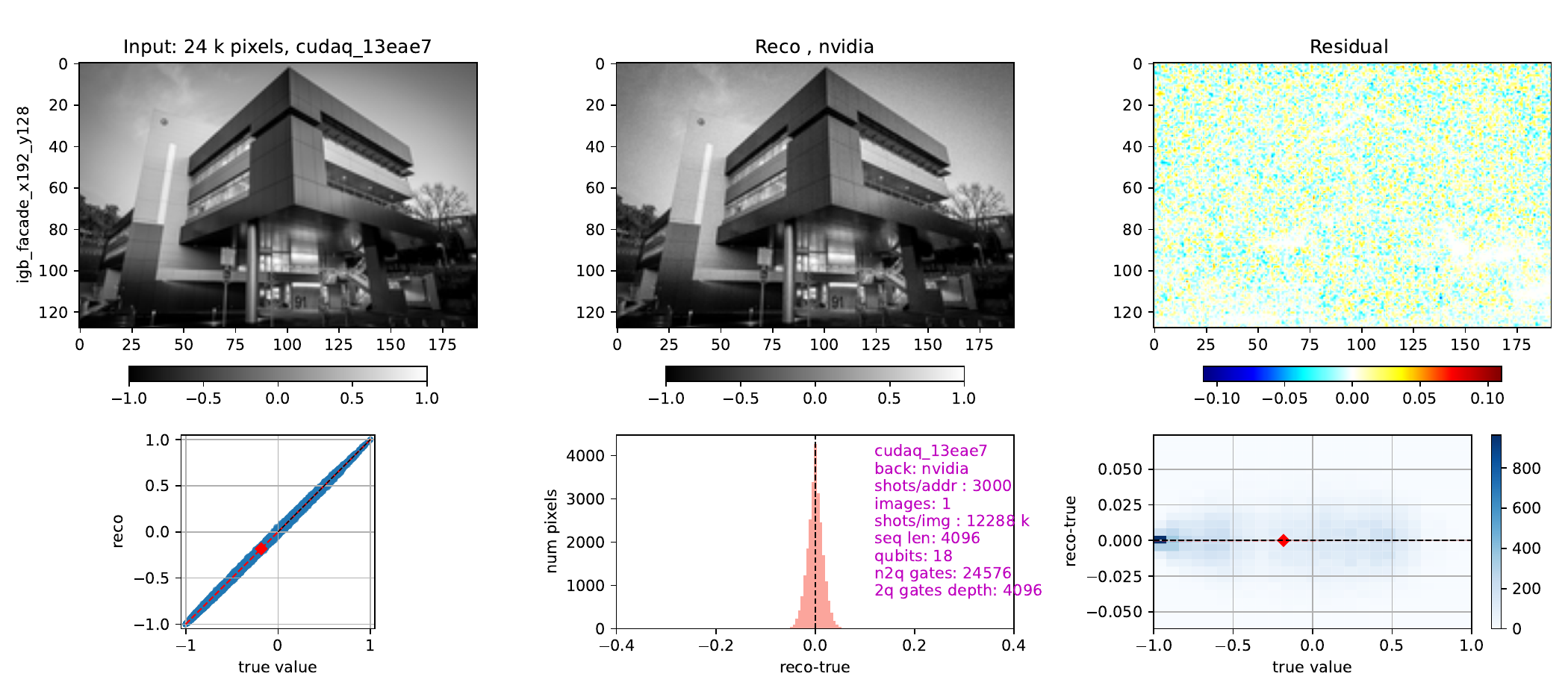}\par
    \vspace*{2pt}%
    \includegraphics[width=\linewidth]{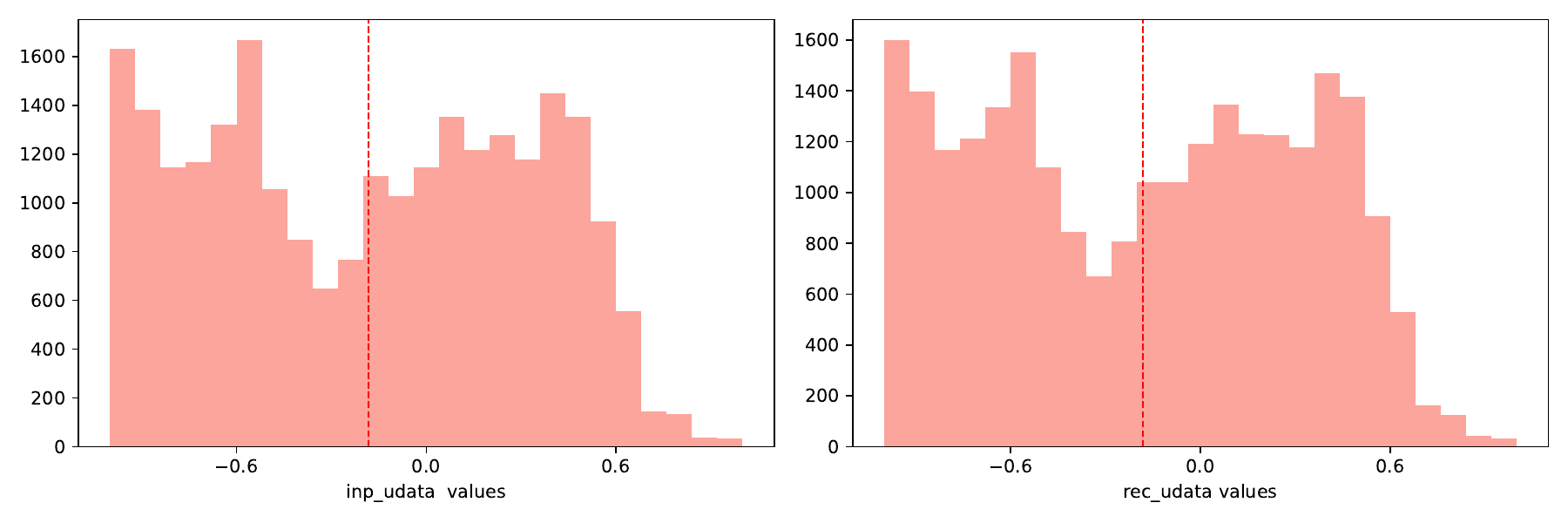}
    \caption{25 k-pixel grayscale image (Building).}
    \label{fig:building_combined}
  \end{subfigure}\hfill
  %
  \begin{subfigure}{.48\textwidth}
    \centering
    \includegraphics[width=\linewidth]{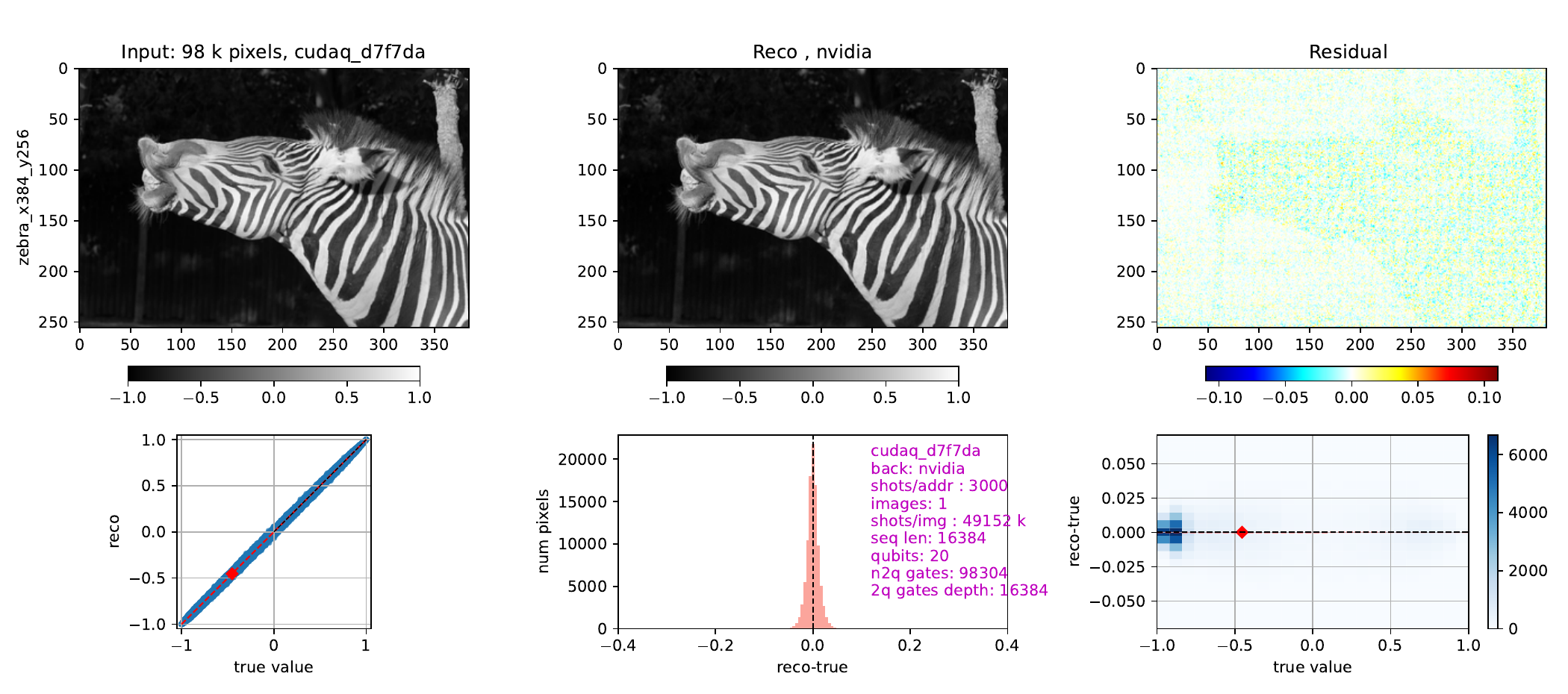}\par
    \vspace*{2pt}%
    \includegraphics[width=\linewidth]{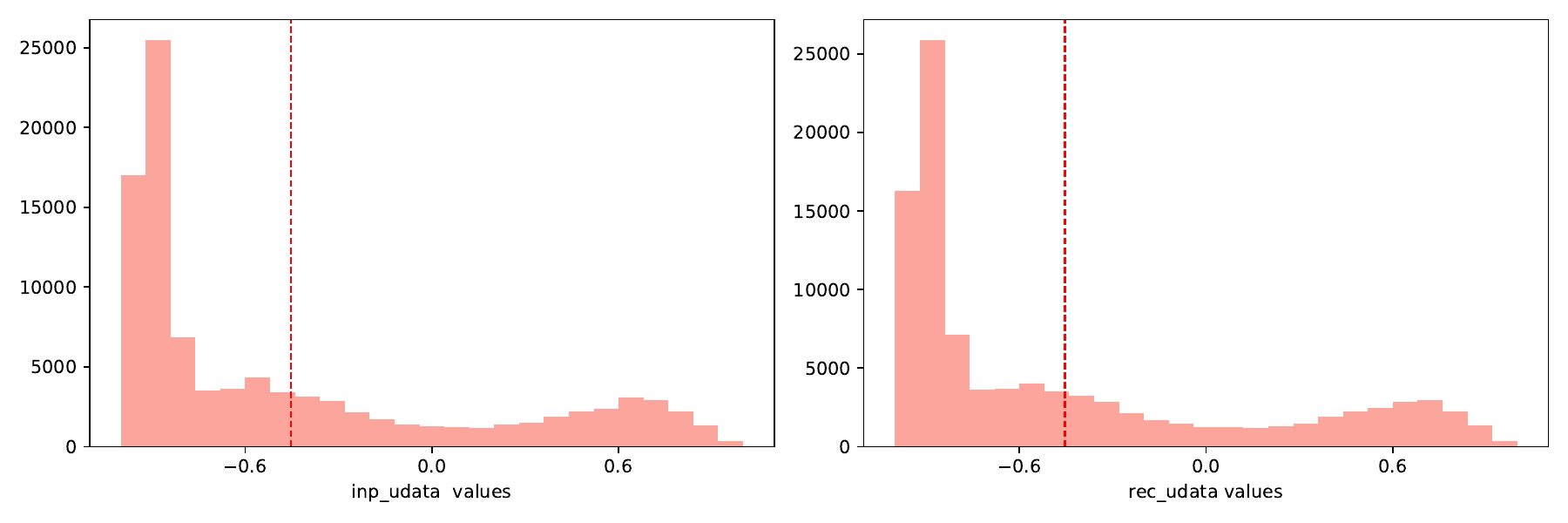}
    \caption{98 k-pixel grayscale image (Zebra).}
    \label{fig:zebra_combined}
  \end{subfigure}

    \caption{
Quantum image–encoding and reconstruction results for four grayscale images of different resolutions.
In each sub-figure the top panel shows the reconstruction correlation, the reconstruction distribution, and the residual encoding error, whereas the bottom panel presents the overall reconstruction benchmark.}
  \label{fig:all_benchmarks}
\end{figure*}

\end{document}